\documentclass[aps,prl,twocolumn,showpacs,groupedaddress]{revtex4}  
\usepackage{graphicx}  
\usepackage{dcolumn}   
\usepackage{bm}        
\usepackage{amssymb}   

\begin{document}


\hspace{5.2in} \mbox{Fermilab-Pub-07/196-E}

\title{Direct observation of the strange $b$ baryon $\Xi_b^-$}
%
\author{V.M.~Abazov$^{35}$}
\author{B.~Abbott$^{75}$}
\author{M.~Abolins$^{65}$}
\author{B.S.~Acharya$^{28}$}
\author{M.~Adams$^{51}$}
\author{T.~Adams$^{49}$}
\author{E.~Aguilo$^{5}$}
\author{S.H.~Ahn$^{30}$}
\author{M.~Ahsan$^{59}$}
\author{G.D.~Alexeev$^{35}$}
% ...

\affiliation{\vspace{0.1 in}(The D\O\ Collaboration)\vspace{0.1 in}}
% ...
\affiliation{$^{5}$University of Alberta, Edmonton, Alberta, Canada,
                Simon Fraser University, Burnaby, British Columbia, Canada,
                York University, Toronto, Ontario, Canada, and
                McGill University, Montreal, Quebec, Canada}
% ...
\affiliation{$^{28}$Tata Institute of Fundamental Research, Mumbai, India}
% ...
\affiliation{$^{30}$Korea Detector Laboratory, Korea University, Seoul, Korea}
% ...
\affiliation{$^{35}$Joint Institute for Nuclear Research, Dubna, Russia}
% ...
\affiliation{$^{49}$Florida State University, Tallahassee, Florida 32306, USA}
% ...
\affiliation{$^{51}$University of Illinois at Chicago,
                Chicago, Illinois 60607, USA}
% ...
\affiliation{$^{59}$Kansas State University, Manhattan, Kansas 66506, USA}
% ...
\affiliation{$^{65}$Michigan State University,
                East Lansing, Michigan 48824, USA}
% ...
\affiliation{$^{75}$University of Oklahoma, Norman, Oklahoma 73019, USA}
% ...
\date{\today}

\begin{abstract}
We report the first direct observation of the strange $b$ baryon $\Xi_b^-\thinspace (\overline{\Xi}_b^+)$. We reconstruct the decay $\Xi_b^-\thinspace  \to J/\psi\thinspace\Xi^-$, with $J/\psi\to\mu^+\mu^-$, and $\Xi^-\to\Lambda\pi^-\to p\pi^-\pi^-$  in $p\bar{p}$ collisions at $\sqrt{s}=1.96$ TeV. Using 1.3 fb$^{-1}$ of data collected by the D0 detector, we observe $15.2\pm 4.4\thinspace ({\rm stat.})_{-0.4}^{+1.9}\thinspace ({\rm syst.})$ $\Xi_b^-$ candidates at a mass of $5.774\pm 0.011\thinspace ({\rm stat.}) \pm 0.015\thinspace ({\rm syst.})$~GeV. The significance of the observed signal is $5.5\sigma$, equivalent to a probability of $3.3\times 10^{-8}$ of it arising from a background fluctuation. Normalizing to the decay $\Lambda_b\to J/\psi\thinspace \Lambda$, we measure the relative rate
$$\frac{\sigma(\Xi_b^-)\times\mathcal{B}(\Xi_b^-\to J/\psi\thinspace\Xi^-)}{\sigma(\Lambda_b)\times\mathcal{B}(\Lambda_b\to J/\psi\thinspace\Lambda)} = 0.28\pm 0.09\thinspace ({\rm stat.})_{-0.08}^{+0.09}\thinspace ({\rm syst.}).$$

\end{abstract}

\maketitle 


The quark model of hadrons~\cite{pdg} predicts the existence of a number of baryons containing $b$ quarks, with a hierarchical structure similar to that of charmed baryons. Despite significant progress in studying $b$ hadrons over the last decade, only the $\Lambda_b\ (udb)$ $b$ baryon has been directly observed. The $\Xi_b^-\ (dsb)$ (charge conjugate states are assumed throughout this Letter) is a strange $b$ baryon made of valence quarks from all three known generations of fermions and is expected to decay through the weak interaction. Theoretical calculations of heavy quark effective theory~\cite{hqet} and nonrelativistic QCD~\cite{nrqcd} predict the $\Xi_b^-$ mass in the range $5.7-5.8$~GeV~\cite{theory}.

Experiments at the CERN LEP $e^+e^-$ collider have reported indirect evidence of the $\Xi_b^-$ baryon based on an excess of same-sign $\Xi^- \ell^-$ events in jets~\cite{lep}. Interpreting the excess as the semi-inclusive $\Xi_b^- \to\Xi^-\ell^-\bar{\nu}_\ell X$ decay,  the average lifetime of the $\Xi_b^-$ is $1.42^{+0.28}_{-0.24}$ ps~\cite{hfavg}. In this Letter, we report the first direct observation of the $\Xi_b^-$ baryon,
fully reconstructed in an exclusive decay. We observe the decay $\Xi_b^-\to J/\psi\thinspace \Xi^-$, with $J/\psi\to\mu^+\mu^-$, $\Xi^-\to\Lambda\pi^-$, and $\Lambda\to p\pi^-$. The analysis is based on a data sample of 1.3 fb$^{-1}$ integrated luminosity collected in $p\bar{p}$ collisions at $\sqrt{s}=1.96$ TeV with the D0 detector at the Fermilab Tevatron collider during $2002-2006$. 

The D0 detector is described in detail elsewhere~\cite{d0det}. The components most relevant to this analysis are the central tracking system and the muon spectrometer. The central tracking system consists of a silicon microstrip tracker (SMT) and a central fiber tracker (CFT) that are surrounded by a 2~T superconducting solenoid. The SMT is optimized for tracking and vertexing for the pseudorapidity region $|\eta|<3$ ($\eta=-\ln[\tan(\theta/2)]$ and $\theta$ is the polar angle) while the CFT has coverage for $|\eta|<2$. Liquid-argon and uranium calorimeters in a central and two end-cap cryostats cover the pseudorapidity region $|\eta|<4.2$. The muon spectrometer is located outside the calorimeter and covers the pseudorapidity region $|\eta|<2$. It comprises a layer of drift tubes and scintillator trigger counters in front of 1.8~T iron toroids followed by two similar layers behind the toroids.

\begin{figure}[htb]
\begin{center}
\includegraphics[width=1.1in]{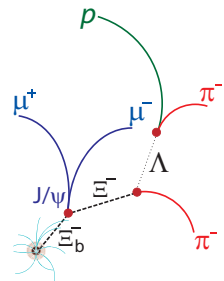}
\end{center}
\caption{Schematic of the $\Xi_b^-\to J/\psi\thinspace\Xi^-\to J/\psi\thinspace \Lambda\pi^-\to (\mu^+\mu^-)\ (p\pi^-)\pi^-$ decay topology. The $\Lambda$ and $\Xi^-$ baryons have decay lengths of the order of cm; the $\Xi_b^-$ has an expected decay length of the order of mm.}
\label{fig:XibSchem}
\end{figure}

The topology of $\Xi_b^-\to J/\psi\thinspace\Xi^-\to J/\psi\thinspace\Lambda\pi^-$ decay (see Fig.~\ref{fig:XibSchem}) is similar to that of the $\Lambda_b\to J/\psi\thinspace\Lambda$ decay; therefore, 
the reconstruction of the $J/\psi$ and $\Lambda$ and their selection discussed below are guided by the strategies applied to the $\Lambda_b$ lifetime measurement in D0~\cite{LbLTime}. They are then validated with simulated Monte Carlo (MC) $\Xi_b^-$ events.  The {\sc pythia} MC program~\cite{pythia} is used to generate $\Xi_b^-$ signal events while the {\sc EvtGen} program~\cite{evtgen} is used to simulate $\Xi_b^-$ decays. The $\Xi_b^-$ mass and lifetime are set to be 5.840~GeV and 1.33~ps respectively, their default values in these programs.
The generated events are subjected to the same reconstruction and selection programs as the data after passing through the D0 detector simulation based on the {\sc geant} package~\cite{geant}. MC events are reweighted using the weights determined by matching transverse momentum ($p_T$) distributions of $J/\psi$, proton and pion from the $\Lambda_b\to J/\psi\thinspace\Lambda\to J/\psi\thinspace p\pi^-$ decays in MC to those observed in the data.

$J/\psi\to\mu^+\mu^-$ decays are reconstructed from two oppositely charged muons that have a common vertex. Muons are identified by matching tracks reconstructed in the central tracking system with either track segments in the muon spectrometer or calorimeter energies consistent with the muon trajectory. They are required to have $p_T>1.5$~GeV and at least one of them must be reconstructed in each of the three muon drift tube layers. The dimuon invariant mass $M(\mu^+\mu^-)$ is required to be in the range $2.5-3.6$~GeV. 
In addition, events must have at least one reconstructed primary vertex of the $p\bar{p}$ interaction. If two or more vertices are reconstructed, the one closest to the reconstructed $\Xi_b^-$ vertex (see below) is used.
Events containing a $J/\psi$ candidate are reprocessed with a version of the track reconstruction algorithm that improves the efficiency for tracks with low $p_T$ and high impact parameters. Consequently, the efficiencies for $K_S^0$, $\Lambda$, and $\Xi^-$ reconstruction are significantly increased. Figure~\ref{fig:XiMass}(a) shows the invariant mass distributions of the reconstructed $\Xi^-$ candidates (see below) before and after the reprocessing. The reprocessing increases the $\Xi^-$ yield by approximately a factor of 5.5. 
For further analysis, $J/\psi\to\mu^+\mu^-$ candidates are required to have mass $2.80<M(\mu^+\mu^-)<3.35$~GeV and $p_T>5$~GeV. The mass windows here and below are
chosen to be approximately $\pm 5\sigma$ and the $p_T$ requirement ensures that
the selected $J/\psi$ candidates are above the sharp turn-on of the detector and trigger acceptances. 

$\Lambda\to p\pi^-$ candidates are formed from two oppositely charged tracks that originate from a common vertex. The track with the higher $p_T$ is assumed to be the proton. MC studies show that this assignment gives nearly 100\% correct combination. The invariant mass of the $p\pi^-$ pair must have a mass between 1.105 and 1.125~GeV. The two tracks are required to have a total of no more than two hits in the tracking detector before the reconstructed $p\pi^-$ vertex. Furthermore, the impact parameter significance (the impact parameter with respect to the event vertex divided by its uncertainty) must exceed three for both tracks and exceed four for at least one of them. These selection cuts are the same as those in Ref.~\cite{LbLTime}.

\begin{figure}[htb]
\begin{center}
    \begin{tabular}{cc}
        \includegraphics[width=1.65in]{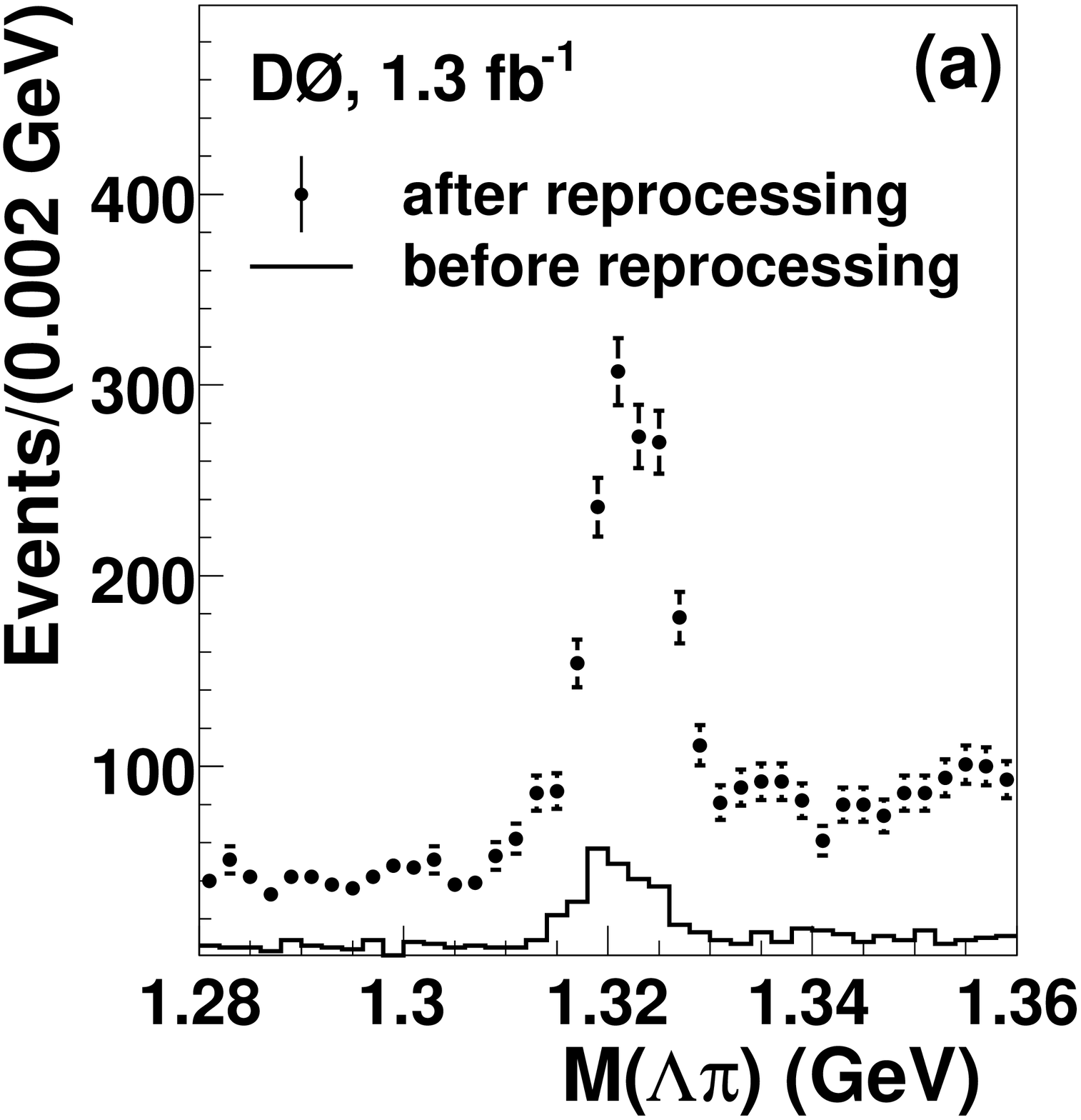} &
        \includegraphics[width=1.65in]{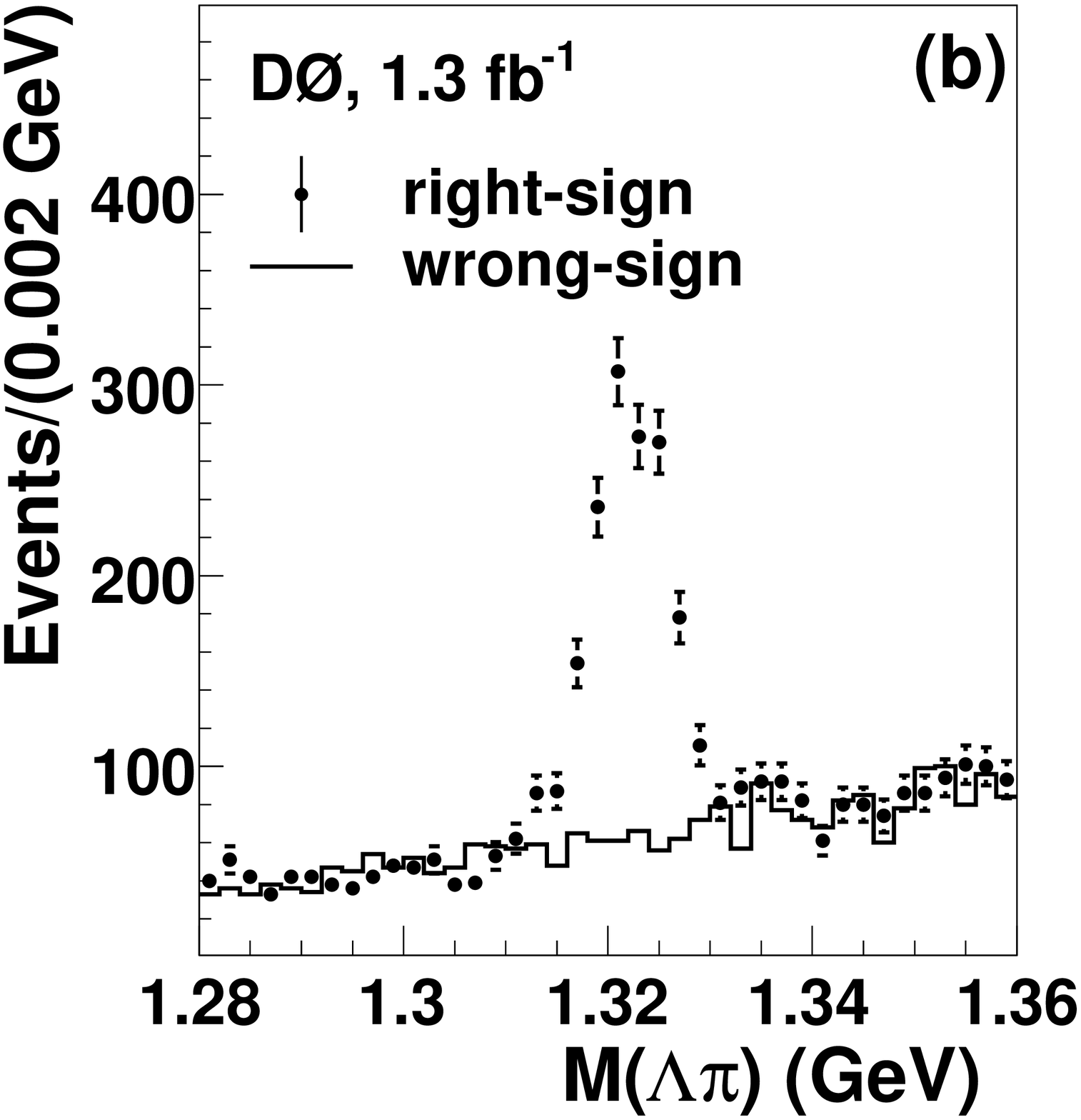}
    \end{tabular}
\end{center}
\caption{Invariant mass distributions of the $\Lambda\pi$ pair before the $\Xi_b^-$ reconstruction for (a) the right-sign $\Lambda\pi^-$ combinations before and after reprocessing and (b) the right-sign $\Lambda\pi^-$ and the wrong-sign $\Lambda\pi^+$ combinations after reprocessing. The reprocessing significantly increases the $\Xi^-$ yield. Fits to the post-reprocessing distributions of the right-sign combination with a Gaussian signal and a first-order polynomial background yield $603\pm 34$ $\Xi^-$'s and $548\pm 31$ $\overline{\Xi}^+$'s. }
\label{fig:XiMass}
\end{figure}

The $\Lambda$ candidates are then combined with negatively charged tracks (assumed to be pions) to form $\Xi^-\to\Lambda\pi^-$ decay candidates. The pion must have an impact parameter significance greater than three.  The $\Lambda$ and the pion are required to have a common vertex. For both $\Lambda$ and $\Xi^-$ candidates, the distance between the event vertex and its decay vertex is required to exceed four times its uncertainty. Moreover, the uncertainty of the distance between the production vertex and its decay vertex (decay length) in the transverse plane (the plane perpendicular to the beam direction) must be less than 0.5 cm. These two requirements reduce combinatoric and track mismeasurement backgrounds.  

The two pions from $\Xi^-\to\Lambda\pi^-\to(p\pi^-)\pi^-$ decays (right-sign) have the same charge. Consequently, the combination $\Lambda\pi^+$ (wrong-sign) events form an ideal control sample for background studies. Figure~\ref{fig:XiMass}(b) compares mass distributions of the right-sign $\Lambda\pi^-$ and the wrong-sign $\Lambda\pi^+$ combinations. The $\Xi^-$ mass peak is evident in the distribution of the right-sign events. A $\Lambda\pi^-$ pair is considered to be a $\Xi^-$ candidate if its mass is within the range $1.305<M(\Lambda\pi^-)<1.340$~GeV.

$\Xi_b^-\to J/\psi\thinspace\Xi^-$ decay candidates are formed from $J/\psi$ and $\Xi^-$ pairs that originate from a common vertex and have an opening angle in the transverse plane less than $\pi/2$ rad. The uncertainty of the proper decay length of the $J/\psi\thinspace\Xi^-$ vertex must be less than 0.05 cm in the transverse plane. A total of 2308 events remains after this preselection.
The wrong-sign events are subjected to the same preselection as the right-sign events. A total of 1124 wrong-sign events is selected as the control sample.

Several distinctive features of the $\Xi_b^-\to J/\psi\thinspace\Xi^-\to J/\psi\thinspace\Lambda\pi^-\to (\mu^+\mu^-)\ (p\pi^-)\pi^-$ decay are utilized to further suppress backgrounds. The wrong-sign background events from the data and MC signal $\Xi_b^-$ events are used for studying additional event selection criteria. Protons and pions from the $\Xi^-$ decays of the $\Xi_b^-$ events are expected to have higher momenta than those from most of the background processes. 
Therefore, protons are required to have $p_T>0.7$~GeV. Similarly, minimum $p_T$ requirements of 0.3 and 0.2~GeV are imposed on pions from $\Lambda$ and $\Xi^-$ decays, respectively. These requirements remove 91.6\% of the wrong-sign background events while keeping 68.7\% of the MC $\Xi_b^-$ signal events. 
Backgrounds from combinatorics and other $b$ hadrons are reduced by using topological decay information. Contamination from decays such as $B^-\to J/\psi\thinspace K^{*-}\to J/\psi\thinspace K_S^0\pi^-$ and $B^0\to J/\psi\thinspace K^{*-}\pi^+\to J/\psi\thinspace (K_S^0\pi^-)\pi^+$ are suppressed by requiring the $\Xi^-$ candidates to have decay lengths greater than 0.5 cm and $\cos(\theta_{\rm col})>0.99$, as the $\Xi^-$ baryons in MC have an average decay length of 4.8~cm. Here $\theta_{\rm col}$ is the angle between the $\Xi^-$ direction and the direction from the $\Xi^-$ production vertex to its decay vertex in the transverse plane. These two requirements on the $\Xi^-$ reduce the background by an additional 56.4\%, while removing only 1.7\% of the MC signal events. The contribution from the $\Omega_b^-$ baryon is estimated to be negligible. 
Finally, $\Xi_b^-$ baryons are expected to have a sizable lifetime. To reduce prompt backgrounds, the transverse proper decay length significance of the $\Xi_b^-$ candidates is required to be greater than two. This final criterion retains 83.1\% of the MC signal events but only 43.9\% of the remaining background events. 

\begin{figure}[htb]
\begin{center}
\mbox{\includegraphics[width=1.65in]{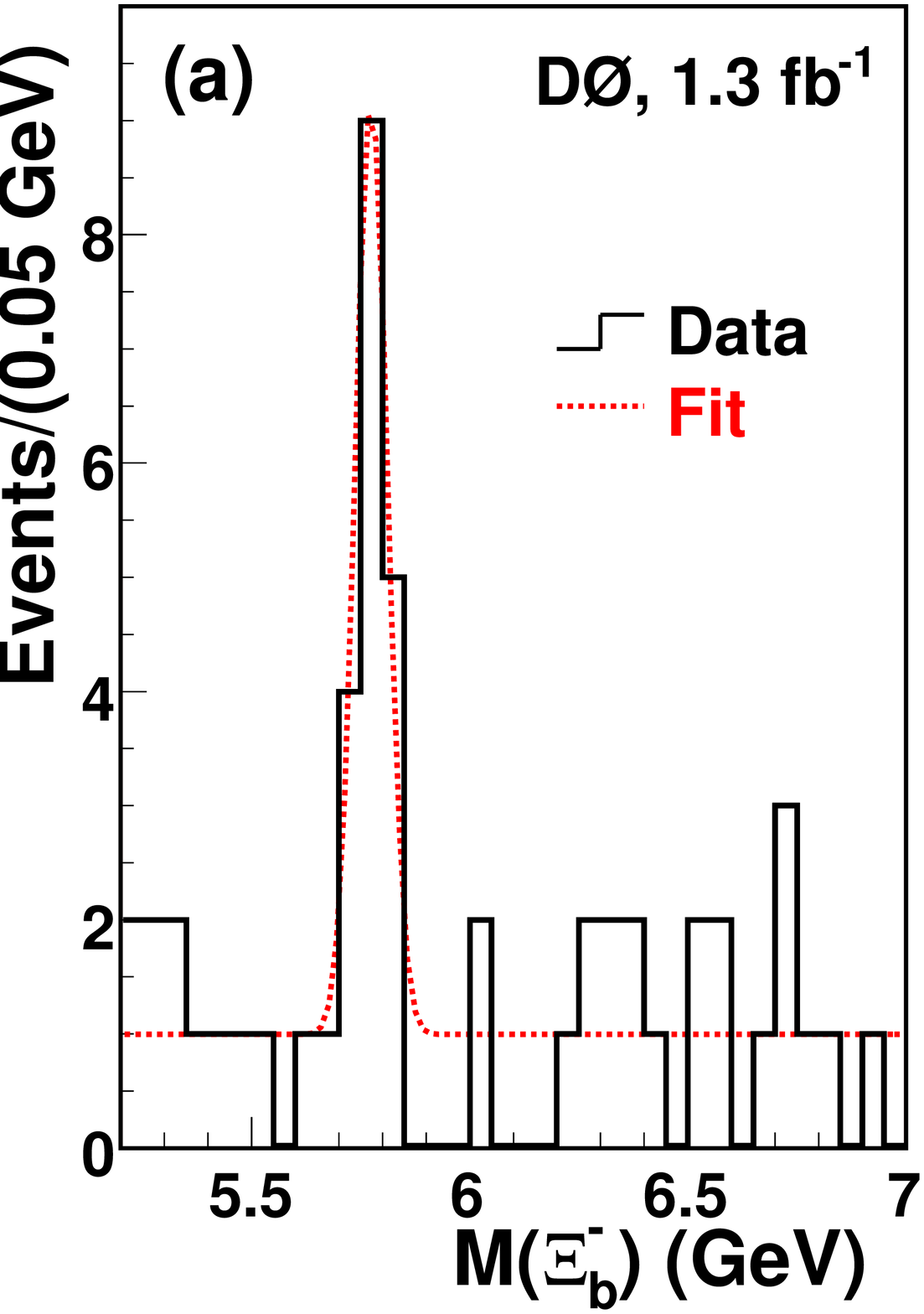}\includegraphics[width=1.65in]{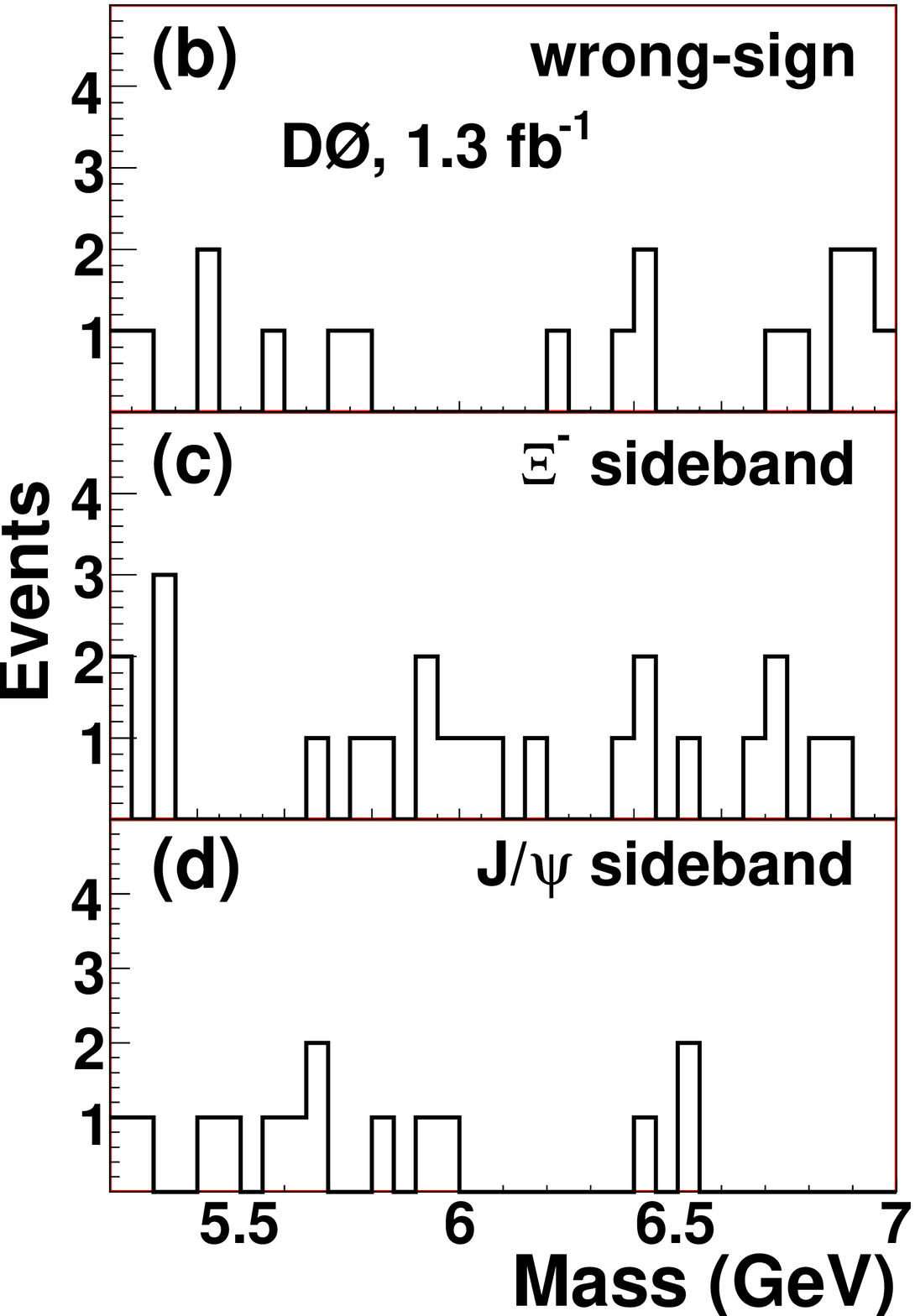}}
\end{center}
\caption{(a) The $M(\Xi_b^-)$ distribution of the $\Xi_b^-$ candidates after all selection criteria. The dotted curve is an unbinned likelihood fit to the model of a constant background plus a Gaussian signal. The $(\mu^+\mu^-)\Lambda\pi$ mass distributions for (b) the wrong-sign background, (c) the $\Xi^-$ sideband, and (d) the $J/\psi$ sideband events. The mass $M(J/\psi\thinspace\Lambda\pi)-M(\mu^+\mu^-)+M_{\rm PDG}(J/\psi)$ is plotted for (b) and (c) while the mass $M(\mu^+\mu^-\Xi^-)-M(\Lambda\pi^-)+M_{\rm PDG}(\Xi^-)$ is plotted for (d).}
\label{fig:XibMass}
\end{figure}

In the data, 51 events with the $\Xi_b^-$ candidate mass between 5.2 and 7.0~GeV pass all selection criteria. The mass range is chosen to be wide enough to encompass masses of all known $b$ hadrons as well as the predicted mass of the $\Xi_b^-$ baryon. The candidate mass, $M(\Xi_b^-)$, is calculated as $M(\Xi_b^-)=M(J/\psi\thinspace\Xi^-)-M(\mu^+\mu^-)-M(\Lambda\pi^-)+M_{\rm PDG}(J/\psi)+M_{\rm PDG}(\Xi^-)$ to improve the resolution. Here $M(J/\psi\thinspace\Xi^-)$, $M(\mu^+\mu^-)$, and $M(\Lambda\pi^-)$ are the reconstructed masses while $M_{\rm PDG}(J/\psi)$ and $M_{\rm PDG}(\Xi^-)$ are taken from Ref.~\cite{pdg}. The distribution of $M(\Xi_b^-)$ is shown in Fig.~\ref{fig:XibMass}(a).  A mass peak near 5.8~GeV is apparent.  A number of cross checks are performed to ensure the observed peak is not due to artifacts of the analysis: 
(1) The $J/\psi\thinspace\Lambda\pi^+$ mass distribution of the wrong-sign events, shown in Fig.~\ref{fig:XibMass}(b), is consistent with a flat background.
(2) The event selection is applied to the sideband events of the $\Xi^-$ mass peak, requiring $1.28< M(\Lambda\pi^-)<1.36$~GeV but excluding the $\Xi^-$ mass window. Similarly, the selection is applied to the $J/\psi$ sideband events with  $2.5<M(\mu^+\mu^-)<2.7$~GeV. The high-mass sideband is not considered due to potential contamination from $\psi'$ events. As shown in Fig.~\ref{fig:XibMass}(c-d), no evidence of a mass peak is present for either $(\mu^+\mu^-)\thinspace(p\pi^-)\pi^-$ distribution.
(3) 
The possibility of a fake signal due to the residual $b$ hadron background is investigated by applying the final $\Xi_b^-$ selection to high statistics MC samples of $B^-\to J/\psi\thinspace K^{*-}\to J/\psi\thinspace K_S^0\pi^-$, $B^0\to J/\psi\thinspace K_S^0$, and $\Lambda_b\to J/\psi\thinspace\Lambda$. No indication of a mass peak is observed in the reconstructed $J/\psi\thinspace\Xi^-$ mass distributions.
(4) The mass distributions of $J/\psi$, $\Xi^-$, and $\Lambda$ are investigated by relaxing the mass requirements on these particles one at a time for events both in the $\Xi_b^-$ signal region and the sidebands. The numbers of these particles determined by fitting their respective mass distribution are fully consistent with the quoted numbers of signal events plus background contributions.
(5) The robustness of the observed mass peak is tested by varying selection criteria within reasonable ranges. All studies confirm the existence of the peak at the same mass.

Interpreting the peak as $\Xi_b^-$ production, candidate masses are fitted with the hypothesis of a signal plus background model using an unbinned likelihood method. The signal and background shapes are assumed to be Gaussian and flat, respectively. The fit results in a $\Xi_b^-$ mass of $5.774\pm 0.011$~GeV with a width of $0.037\pm 0.008$~GeV and a yield of $15.2\pm 4.4$ events. Unless specified, all uncertainties are statistical. Following the same procedure, a fit to the MC $\Xi_b^-$ events yields a mass of $5.839\pm 0.003$~GeV, in good agreement with the 5.840~GeV input mass.  The fitted width of the MC mass distribution is $0.035\pm 0.002$~GeV, consistent with the 0.037~GeV obtained from the data. Since the intrinsic decay width of the $\Xi_b^-$ baryon in the MC is negligible, the width of the mass distribution is thus dominated by the detector resolution. 
To assess the significance of the signal, the likelihood, ${\cal L}_{s+b}$, of the signal plus background fit above is first determined.  The fit is then repeated using the background-only model, and a new likelihood ${\cal L}_b$ is found. The logarithmic likelihood ratio $\sqrt{2\ln({\cal L}_{s+b}/{\cal L}_b)}$ indicates a statistical significance of 5.5$\sigma$, corresponding to a probability of $3.3\times 10^{-8}$ from background fluctuation for observing a signal that is equal to or more significant than what is seen in the data. Including systematic effects from the mass range, signal and background models, and the track momentum scale results in a minimum significance of $5.3\sigma$ and a $\Xi_b^-$ yield of $15.2\pm 4.4\thinspace({\rm stat.})_{-0.4}^{+1.9}\thinspace({\rm syst.})$. 
The significance can also be estimated from the numbers of candidate events and estimated background events. In the mass region of 2.5 times the fitted width centered on the fitted mass, 19 candidate events (8 $J/\psi\thinspace\Xi^-$ and  11 $J/\psi\thinspace\overline{\Xi}^+$) are observed while $14.8\pm 4.3\thinspace ({\rm stat.})_{-0.4}^{+1.9}\thinspace ({\rm syst.})$ signal and $3.6\pm 0.6\thinspace ({\rm stat.})_{-1.9}^{+0.4}\thinspace({\rm syst.})$ background events are estimated from the fit. The probability of backgrounds fluctuating to 19 or more events is $2.2\times 10^{-7}$, equivalent to a Gaussian significance of $5.2\sigma$. 

\begin{figure}[htb]
\begin{center}
\includegraphics[width=2.4in]{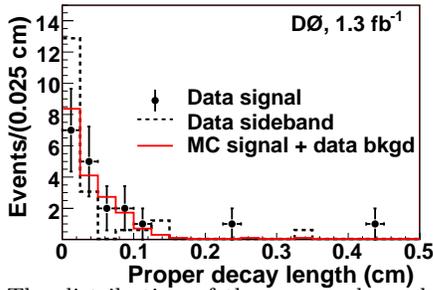}
\end{center}
\caption{The distribution of the proper decay length in the transverse plane of the 19 candidate events in the $\pm 2.5\sigma$ signal mass window along with that of the events in the sidebands, defined to be $5\sigma$ away from the fitted mass. Also shown is the expected distribution from 14.8 MC $\Xi_b^-$ signal events plus 3.6 background events. The distribution of the sideband events is scaled to the number of events in the signal mass window. Kolmogorov-Smirnov tests indicate that the distribution of the signal events is favored over that of the sideband events with respect to the MC expectation by a ratio of five to one.}
\label{fig:XibLTime}
\end{figure}

Figure~\ref{fig:XibLTime} shows distributions of the proper decay length for the 19 candidate events, the $\Xi_b^-$ sideband events, and the MC $\Xi_b^-$ signal events plus estimated background events. The distribution of the candidate events agrees well with that expected from the $\Xi_b^-$ signal while the sideband events have a lower mean proper decay length. Due to the use of lifetime information in the event selection, a $\Xi_b^-$ lifetime measurement is not made in this Letter.

Potential systematic biases on the measured $\Xi_b^-$ mass are studied for the event selection, signal and background models, and the track momentum scale. 
Varying cut values and using a multivariate technique of different variables for event selection leads to a maximum change of 0.020~GeV in the $\Xi_b^-$ mass. Subtracting an estimated statistical contribution to the change, a conservative $\pm 0.015$~GeV systematic uncertainty is assigned due to the event selection.
Using double Gaussians for the signal model, a first-order polynomial for the background model, or fixing the  mass resolution to that obtained from the MC $\Xi_b^-$ events all lead to negligible changes in the mass. The mass, calculated using the world average values~\cite{pdg} of intermediate particle masses above, is found to have a weak dependence on the track momentum scale. This has been verified using the $\Lambda_b\to J/\psi\thinspace\Lambda$ and $B^0\to J/\psi\thinspace K_S^0$ events observed in the data. A systematic uncertainty of $\pm 0.002$~GeV is assigned, corresponding to the mass difference between our measurement and the world average~\cite{pdg} for the $\Lambda_b$ and $B^0$ hadrons.
Adding in quadrature,  a total systematic uncertainty of $\pm 0.015$~GeV is obtained to yield the measured $\Xi_b^-$mass: $5.774\pm 0.011\thinspace({\rm stat.})\pm 0.015\thinspace({\rm syst.})$~GeV.

The $\Xi_b^-$ $\sigma\times\mathcal{B}$ relative to that of the $\Lambda_{b}$ baryon is calculated using 
$$
\frac{\sigma(\Xi_b^-)\times\mathcal{B}(\Xi_b^-\to J/\psi\thinspace\Xi^-) } {\sigma(\Lambda_b)\times\mathcal{B}(\Lambda_{b} \to J/\psi\thinspace \Lambda)}
=\frac{\epsilon(\Lambda_{b} \to J/\psi\thinspace\Lambda)}{\epsilon(\Xi_b^-\to J/\psi\thinspace\Xi^-)}\frac{N_{\Xi_b^-}}{N_{\Lambda_{b}}}
$$
where $N_{\Xi_b^-}$ and $N_{\Lambda_{b}}$ are the numbers of $\Xi_b^-$ and $\Lambda_{b}$ events reconstructed in data.
Analyzing the same data and using the similar event selection criteria and fitting procedure as the $\Xi_b^-$ analysis, a yield of $240\pm 30\thinspace ({\rm stat.})\pm 12\thinspace({\rm syst.})$ $\Lambda_b$ baryons is determined. 
The efficiencies to reconstruct the decays, $\epsilon(\Xi_b^-)$ and $\epsilon(\Lambda_{b})$, are determined by MC simulation, and the efficiency ratio, $\epsilon(\Lambda_b)/\epsilon(\Xi_b^-)$, is found to be $4.4\pm 1.3$. The uncertainty on $\epsilon(\Lambda_b)/\epsilon(\Xi_b^-)$ arises from MC modeling (27\%), MC statistics (10\%), the reconstruction of the additional pion in the $\Xi_b^-$ decay (7\%), and the $\Xi_b^-$ mass difference between data and MC (5\%). The largest component, MC modeling uncertainty, is due to the difference in the efficiency ratio with and without MC reweighting.  The efficiency ratio is found to be insensitive to changes in $\Lambda_b$ and $\Xi_b^-$ production models. Many other systematic uncertainties on the efficiencies themselves tend to cancel in the ratio of the efficiencies.
We find a relative production ratio of $0.28\pm 0.09\thinspace ({\rm stat.})_{-0.08}^{+0.09}\thinspace ({\rm syst.})$.

In summary, in 1.3 fb$^{-1}$  of data collected by the D0 experiment in $p\bar{p}$ collisions at $\sqrt{s}=1.96$ TeV at the Fermilab Tevatron collider, we have made the first direct observation of the strange $b$ baryon $\Xi_b^-$ with a statistical significance of $5.5\sigma$. We observe the decay mode $\Xi_b^-\thinspace\to J/\psi\thinspace\Xi^-\thinspace $ with $J/\psi\to\mu^+\mu^-$, $\Xi^- \to\Lambda\pi^-\to p\pi^-\pi^-$.  We measure the $\Xi_b^-$ mass to be $5.774\pm 0.011\thinspace ({\rm stat.})\pm 0.015\thinspace ({\rm syst.})$~GeV and determine its $\sigma\times\mathcal{B}$ relative to that of the $\Lambda_b$ to be
$0.28\pm 0.09\thinspace ({\rm stat.})_{-0.08}^{+0.09}\thinspace ({\rm syst.}).$

%
We thank the staffs at Fermilab and collaborating institutions, 
and acknowledge support from the 
DOE and NSF (USA);
CEA and CNRS/IN2P3 (France);
FASI, Rosatom and RFBR (Russia);
CAPES, CNPq, FAPERJ, FAPESP and FUNDUNESP (Brazil);
DAE and DST (India);
Colciencias (Colombia);
CONACyT (Mexico);
KRF and KOSEF (Korea);
CONICET and UBACyT (Argentina);
FOM (The Netherlands);
Science and Technology Facilities Council (United Kingdom);
MSMT and GACR (Czech Republic);
CRC Program, CFI, NSERC and WestGrid Project (Canada);
BMBF and DFG (Germany);
SFI (Ireland);
The Swedish Research Council (Sweden);
CAS and CNSF (China);
Alexander von Humboldt Foundation;
and the Marie Curie Program.
%

\end{document}